\documentclass[conference]{IEEEtran}
\IEEEoverridecommandlockouts
\usepackage{cite}
\usepackage{amsmath,amssymb,amsfonts}
\usepackage{algorithmic}
\usepackage{graphicx}
\usepackage{textcomp}
\usepackage{xcolor}
 
\usepackage[breakall]{truncate}
\usepackage{hyperref}  
\usepackage{arydshln}
\usepackage{multirow}
\usepackage{stackengine}

\def\BibTeX{{\rm B\kern-.05em{\sc i\kern-.025em b}\kern-.08em
    T\kern-.1667em\lower.7ex\hbox{E}\kern-.125emX}}

\newcommand\tmp[1]{#1}  
\newcommand\tp[1]{#1}  

\newcommand\first[1]{\textbf{#1}}  
\newcommand\second[1]{\underline{\smash{#1}}}  
\newcommand\example[1]{\textit{#1}}

\begin{document}

\title{\tmp{
Multitask learning in Audio Captioning: a sentence embedding regression loss acts as a regularizer
}}

\author{
    \IEEEauthorblockN{Etienne Labbé}
    \IEEEauthorblockA{\textit{IRIT, Université de Toulouse} \\
    CNRS, UT3, Toulouse, France \\
    etienne.labbe@irit.fr}
    \and
    \IEEEauthorblockN{Julien Pinquier}
    \IEEEauthorblockA{\textit{IRIT, Université de Toulouse} \\
    CNRS, UT3, Toulouse, France \\
    julien.pinquier@irit.fr}
    \and
    \IEEEauthorblockN{Thomas Pellegrini}
    \IEEEauthorblockA{\textit{IRIT, Université de Toulouse} \\
    CNRS, UT3, Toulouse, France \\
    thomas.pellegrini@irit.fr}
}

\maketitle


\begin{abstract}
\tmp{
In this work, we propose to study the performance of a model trained with a sentence embedding regression loss component for the Automated Audio Captioning task. This task aims to build systems that can describe audio content with a single sentence written in natural language. Most systems are trained with the standard Cross-Entropy loss, which does not take into account the semantic closeness of the sentence. We found that adding a sentence embedding loss term reduces overfitting, but also increased SPIDEr from 0.397 to 0.418 in our first setting on the AudioCaps corpus. When we increased the weight decay value, we found our model to be much closer to the current state-of-the-art methods, with a SPIDEr score up to 0.444 compared to a 0.475 score. Moreover, this model uses eight times less trainable parameters. In this training setting, the sentence embedding loss has no more impact on the model performance.
}
\end{abstract}

\begin{IEEEkeywords}
sound event description, multitask learning, audio language task, overfitting, sentence embedding regression loss, semantic loss
\end{IEEEkeywords}

\section{Introduction}
In recent years, new machine learning systems have been significantly improved for text processing, generation, and understanding, leading to the use of natural language as a global interface between humans and machines. Free-form text can contain much more information than a predefined set of classes, which could improve the machine understanding of our world. In audio, most of the tasks are focused on classification and localization of sound events. Following this idea, the Automated Audio Captioning (AAC) task appeared in 2017~\cite{drossos:2017:waspaa} and aims to create systems that generate a sentence written in natural language that describes an audio file. The audio can contain various sound events (human, natural, domestic, urban, music, effects...) of different lengths, recorded with different devices and in different scenes. The description can contain any kind of detail in the audio, with temporal or spatial relations between them (followed by, in the background...) or different characterizations (high-pitched, short, repetitive...). Since the descriptions are written by humans, we need to consider different words used to describe similar sounds (\example{Birds are calling} / \example{chirping} / \example{singing} / \example{tweeting}), different sentence structures (\example{A door that needs to be oiled} / \example{A door with squeaky hinges}), subjectivity (\example{Man speaks in a foreign language}), high-level descriptions (\example{A vulgar man speaks} / \example{Unintelligible conversation}), and vagueness (\example{Someone speaks} instead of \example{A man gives a speech over a reverberating microphone}).

In AAC, most approaches use deep learning models trained with the standard Cross-Entropy (CE) loss. However, this loss tends to \tp{generate} repetitive and generic content~\cite{https://doi.org/10.48550/arxiv.1904.09751} and does not take into account synonyms, various sentences structures or the semantic closeness. Several studies introduced another criterion, the Self-Critical Sequence Training~\cite{https://doi.org/10.48550/arxiv.1612.00563} (SCST) used in reinforcement learning to fine-tune the model directly on a metric instead of the loss. This technique relies on sampling the next word to generate a new sentence. If this sentence has a higher score than the original one, the model is rewarded and the outputs probabilities for this new sentence are encouraged. However, this technique leads to degenerated sentences~\cite{https://doi.org/10.48550/arxiv.2108.02752}, with repetitive n-grams without syntactical correctness. 



Motivated by the limitations of CE and SCST, in this work, we attempted to add a Sentence Embedding Regression (SER) loss used in~\cite{sent_emb_reg_loss} to improve our model. \tmp{We begin this paper by describing our baseline system and then explain how to add SER loss. We present related work in which we compare and then describe the detailed hyperparameters. Finally, we present the results and discuss the differences. 
}

\section{Baseline system description}

We use an encoder-decoder architecture widely used in AAC systems, with an encoder pre-trained on AudioSet~\cite{gemmeke_audio_2017} to extract a strong representation of sounds events. More specifically, we used the \texttt{CNN14\_DecisionLevel\_Att} audio encoder from the Pre-trained Audio Neural Networks study (PANN)~\cite{kong_panns_2020}, with the pre-trained weights available on Zenodo\footnote{\url{https://zenodo.org/record/3987831}}. This architecture gives the best results on the classification of sound events in the audio captioning dataset part when compared to the other PANN architectures available. We have found that freezing weights does not decrease performance while significantly speeding up the training process. This encoder provides sequences of embeddings of dimension $31 \times 2048$ for ten-second long audio recordings. On top of that, we add a projection layer to get 256-dimensional embeddings to match the input dimension of the decoder $d_{model}$.

The decoder is a standard transformer decoder~\cite{https://doi.org/10.48550/arxiv.1706.03762} with 6 layers, 4 attention heads per layer, a global embedding size $d_{model}$ set to 256 and a global dropout probability of 0.2. We also used the GELU~\cite{https://doi.org/10.48550/arxiv.1606.08415} activation layer in the decoder.

The decoder is trained using teacher forcing, \tp{\textit{i.e.}} gives the ground truth previous reference tokens to the model to predict the next one. The baseline criterion is the standard CE loss over the whole sequence between the output probabilities and the reference token classes.

During inference, we used the beam search algorithm with a beam size set to 2 since higher value does not bring improvements. We conditioned the sentence generation to improve performance and overall caption quality: by limiting the prediction length to a minimum of 3 tokens and a maximum of 30 tokens and by forbidding the model to generate the same token twice, except for stop-word tokens predefined in the Natural Language ToolKit (NLTK)~\cite{bird2009natural} package. These constraints reduce the number of invalid sentences and repetitions and give a slight improvement in the performance of our model.

\section{Adding a sentence embedding regression loss}

\subsection{Sentence-BERT model}
The Sentence-BERT~\cite{https://doi.org/10.48550/arxiv.1908.10084} (SBERT) model is a transformer-based model which combines a BERT~\cite{DBLP:journals/corr/abs-1810-04805} model with a pooling and a projection layer to produce a single embedding of a fixed size of 768 values for a given sentence.

\tp{Available} SBERT models \tp{have} been trained on two text databases: the Stanford Natural Language Inference (SNLI)~\cite{bowman-etal-2015-large} and the Multi-Genre Natural Language Inference (MultiNLI)~\cite{williams-etal-2018-broad}. These datasets contain pairs of two sentences annotated with contradiction, entailment or neutral label. To learn sentence semantic, two SBERT embedding are fed to a classification layer, which must predicts the label of the pair.

\subsection{SER loss}
To use the SBERT model to improve our model, we need to use the same token units to have the same sequence size. BERT uses WordPiece tokens~\cite{https://doi.org/10.48550/arxiv.1609.08144} instead of words which form a vocabulary of 30522 different units in our experiments.

\begin{figure}[htbp]
    \centering
    \includegraphics[width=0.65\linewidth]{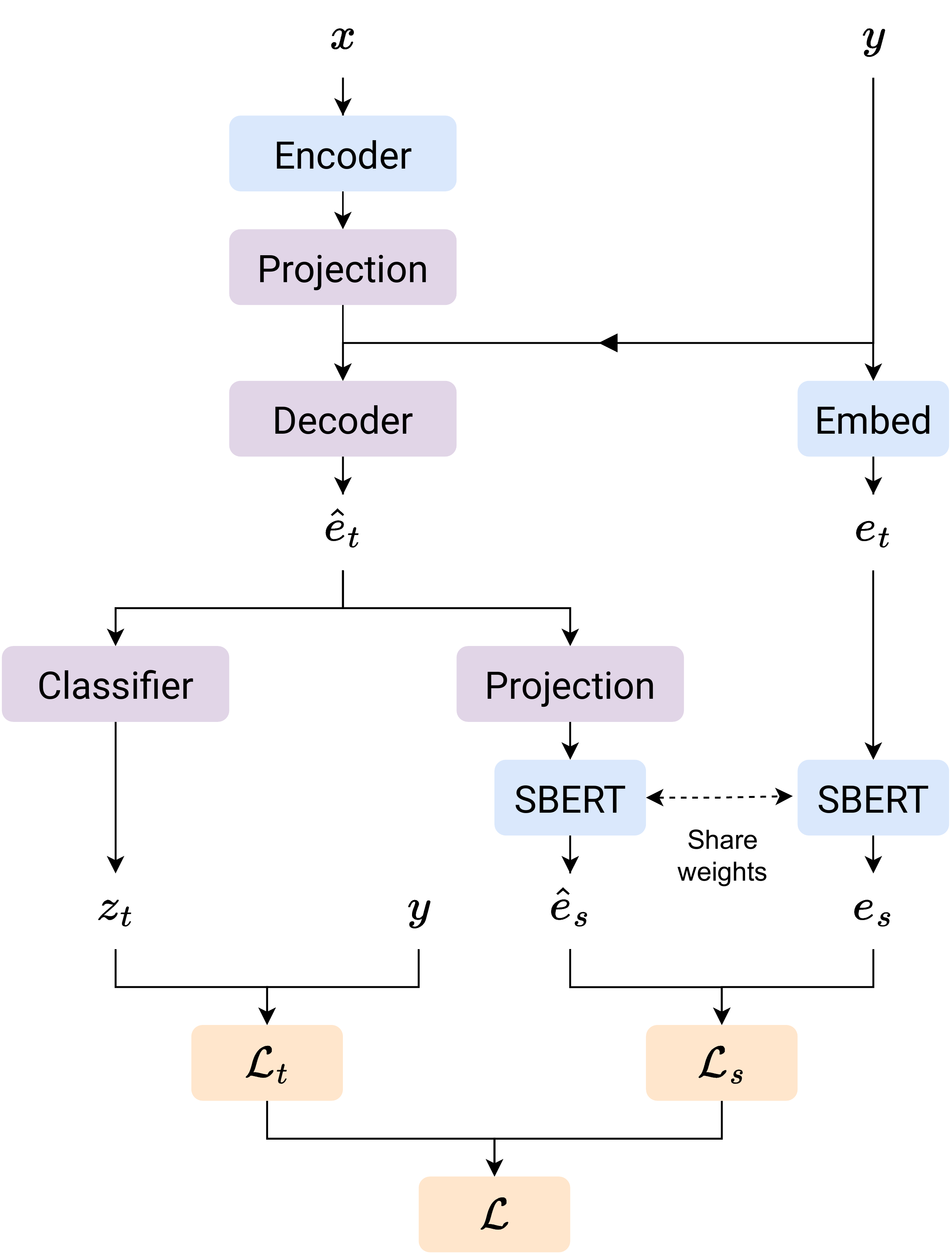}
    \caption{Overview of our proposed training method. The blue boxes are the pre-trained frozen layers, the purple ones the trainable layers and the orange ones the functions. The SBERT block contains the SBERT model with its first embedding layer removed.}
    \label{fig_sbert_loss}
\end{figure}

Fig.~\ref{fig_sbert_loss} resumes the whole procedure and layers used. During training phase, we use audio features and the ground truth previous tokens to generate the next token embeddings named $\hat{e}_t$. These embeddings are used in two different parts of the model. First, they are projected to logits using a classifier for the standard CE loss $\mathcal{L}_t$. The token embeddings are also projected from 256 to 768-dimensional embedding to match the SBERT embedding input shape. The resulting embedding $\hat{e}_s$ is used as input with the ground truth embedding for the SER loss component $\mathcal{L}_s$. In order to use the SBERT model to train our model, we need to remove the first layer of SBERT which maps tokens IDs to embedding vectors (named "Embed" in the figure), since this layer is not differentiable. At inference time, only the classifier branch is used.

We tried several regression criteria for the $\mathcal{L}_{s}$ loss: CosineEmbeddingLoss, L1Loss, MSELoss and SmoothL1Loss. The best one that we obtained is the SmoothL1Loss~\cite{DBLP:journals/corr/Girshick15}, a regression function which combines MSE and L1Loss, described in equation~\eqref{eq_ser_loss}:

\begin{equation}
    \mathcal{L}_{s} (\hat{e}_{s}, e_{s}) =
    \begin{cases}
        (\hat{e}_{s} - e_{s})^2 \cdot \frac{1}{2\beta} & \text{if } |\hat{e}_{s} - e_{s}| < \beta \\
        |\hat{e}_{s} - e_{s}| - \frac{\beta}{2} & \text{otherwise}
    \end{cases}
    \label{eq_ser_loss}
\end{equation}

The $\beta$ hyperparameter control whether MSE or L1Loss must be used. We kept the standard CE as our first component $\mathcal{L}_t$ to help the model to produce syntactically valid sentences. The final loss is given by equation~\eqref{eq_final_sbert_loss} and sum $\mathcal{L}_t$ and $\mathcal{L}_s$, weighted by a coefficient $\lambda$:

\begin{equation}
    \mathcal{L} = \mathcal{L}_{t} + \lambda \cdot \mathcal{L}_{s}
    \label{eq_final_sbert_loss}
\end{equation}

\section{Related work}

The current state-of-the-art on AudioCaps~\cite{https://doi.org/10.48550/arxiv.2210.17143} is a full transformer architecture named Audio Captioning Transformer (ACT)~\cite{act}, pre-trained on AudioSet like PANN models. The system uses mixup~\cite{zhang2018mixup} to improve generalization during training between audio waveform and spectrograms and concatenate the corresponding captions. For inference, Gaussian noise and SpecAugment~\cite{park19e_interspeech} are used to produce several variants of the same sample and are given to the model. The intermediate representations of the same example are averaged to produce a better sentence.

In~\cite{gontier:hal-03522488}, the authors proposed to use a pre-trained transformer decoder named BART~\cite{bart_model} to generate better sentences. Their first encoder (YAMNet) predicts the names of the predicted AudioSet classes to improve the audio representation. These classes names are the inputs of the BART embedding layer and are added to the PANN encoder audio embeddings. The decoder is the pretrained BART transformer decoder part. We named their model BYP in the table for shortening BART+YAMNet+PANNs.

An approach similar to ours has been proposed in~\cite{sent_emb_reg_loss}. There are notable differences (the audio encoder, optimizer and hyperparameters for optimization and generation) which provide a stronger baseline. \tmp{Unlike them, we train our model with only one phase (CE+SER losses) instead of two (CE then CE+SER losses).}

\begin{table*}[ht]
    \centering
    \caption{AAC results on AudioCaps \textbf{testing} subset with captioning metrics. The arrow $\uparrow$ indicates that a higher value in the column is better, while $\downarrow$ indicates that a lower value is better. Optim. and Wd. stand for Optimizer and Weight decay, respectively.}

    \begin{tabular}{|l|c|c|c|c|c|c|c|c|c|c|c|}
        \hline
        \multirow{2}{*}{\textbf{Method}} & \multirow{2}{*}{\textbf{Optim.}} & \multirow{2}{*}{\textbf{Wd.}} & \textbf{CIDEr-D} & \textbf{SPICE} & \textbf{SPIDEr} & \textbf{FENSE} & \textbf{SBERT} & \textbf{FluErr} & \textbf{\#Words} & \textbf{Trainable} & \textbf{Frozen} \\
        &&& $\uparrow$ & $\uparrow$ & $\uparrow$ & $\uparrow$ & $\uparrow$ & $\downarrow$ & $\uparrow$ & \textbf{params} & \textbf{params} \\
        \hline
        Cross-referencing & N/A & N/A & {.901} & {.217} & {.559} & {.680} & {.682} & {.005} & {952.2} & 0 & 0 \\
        \hline   
        Multi-TTA~\cite{https://doi.org/10.48550/arxiv.2210.17143} & \multirow{2}{*}{AdamW} & $10^{-6}$ & \first{.769} & \first{.181} & \first{.475} & \multirow{2}{*}{N/A} & \multirow{2}{*}{N/A} & \multirow{2}{*}{N/A} & \multirow{2}{*}{N/A} & 108M & \multirow{2}{*}{0} \\
        BYP~\cite{gontier:hal-03522488} & & N/A & \second{.753} & \second{.176} & \second{.465} & & & & & 408M & \\
        \hdashline
        CNN10-trans\cite{sent_emb_reg_loss} & \multirow{2}{*}{Adam} & \multirow{2}{*}{N/A} & {.573} & {.158} & {.365} & \multirow{2}{*}{N/A} & {.545} & \multirow{2}{*}{N/A} & \multirow{2}{*}{N/A} & \multirow{2}{*}{14M} & 0 \\
        +SER cosine loss & & & {.573} & {.166} & {.370} & & {.555} & & & & N/A \\
        \hline
        Our baseline & & & {.628} & {.165} & {.397} & {.595} & {.601} & {.034} & \first{485.8} & 12.4M & 79.7M \\
        +SBERT tokens & AdamW & $10^{-6}$ & {.659} & {.167} & {.413} & {.602} & {.607} & {.034} & {445.0} & 25.7M & 79.7M \\
        +SER loss & & & {.665} & {.170} & {.418} & {.607} & {.614} & {.027} & \second{465.6} & 25.9M & 146.7M \\
        \hdashline
        Our baseline & & & \second{.712} & \first{.176} & \second{.444} & \second{.619} & \first{.621} & \second{.004} & {387.2} & 12.4M & 79.7M \\
        +SBERT tokens & AdamW & $2$ & \first{.715} & \second{.175} & \first{.445} & \first{.620} & \first{.621} & \first{.002} & {390.2} & 25.7M & 79.7M \\
        +SER loss & & & \first{.715} & {.172} & {.443} & \second{.619} & \second{.620} & \first{.002} & {348.8} & 25.9M & 146.7M \\
        \hline
    \end{tabular}
    \label{tab_results}
\end{table*}

\section{Experimental setup}

\subsection{Dataset}
We train and evaluate our models on the AudioCaps~\cite{kim-etal-2019-audiocaps} dataset, which is the largest known audio-language dataset with human generated captions. The audio files are 10-second clips from AudioSet~\cite{gemmeke_audio_2017} and are extracted from YouTube videos. Since some of the original videos are removed or unavailable, our version of the dataset contains 46230 over 49838 files in training subset, 464 over 495 in validation subset and 912 over 975 files in testing subset. Each audio is described by one caption in the training subset and five captions in the validation and testing subsets. 

\subsection{Metrics}
\tmp{
We focused only on the captioning metrics and decided to dismiss the translation metrics (BLEU, ROUGE, METEOR) which are mainly based on n-gram overlapping. CIDEr-D~\cite{vedantam_cider_2015} computes the cosine similarity of the TF-IDF scores for common n-grams in candidates and references. SPICE~\cite{anderson_spice_2016} computes the F1-score of the graph edges representing the semantic propositions extracted from the sentences using a parser and grammar rules. SPIDEr~\cite{liu_improved_2017} is the average of CIDEr-D and SPICE and mainly used to rank AAC systems. Since we are also studying a sentence similarity loss, we decided to add three model-based metrics from~\cite{zhou_can_2022}: SBERT, FluErr and FENSE. The SBERT metric correspond to the cosine similarity of the sentence embedding extracted using a SBERT model. FluErr is the fluency error rate detected by a model trained to detect common errors made by captioning systems like incomplete sentence, repeated event, repeated adverb, missing conjunction and missing verb. The FENSE metric is the SBERT score for each sentence, unless an error in the FluErr metric is detected, then it will be divided by 10. Finally, the last metric "\#Words" is the number of unique words used in the candidates in the whole subset.
}

\subsection{Hyperparameters}

Hyperparameters are crucial for training deep learning systems. We found that the model can obtain drastically different scores when trained using different sets of hyperparameters. \tmp{We optimized our hyperparameters to maximize the FENSE score on the validation subset of AudioCaps.} We train our model for a total of 100 epochs $K$ and with a batch size of 512 samples on a single GPU. \tmp{We used the AdamW optimizer~\cite{adamw} with a weight decay (wd) of $10^{-6}$ in the first experiments and set to $2$ to limit overfitting in the second setting. The weight decay is not applied to the bias weights of the network.} We also denote that the network does not converge when using the standard Adam~\cite{kingma_adam_2017} optimizer with a large wd. The initial learning rate $\text{lr}_0$ is set to $5 \cdot 10^{-4}$ at the beginning of the training, and the values of $\beta_1$ and $\beta_2$ are respectively set to 0.9 and 0.999. \tmp{We used cosine scheduler decay updated at the end of each epoch $k$ with the following rule: $\text{lr}_k = \frac{1}{2} \big(1 + \cos ( \frac{k \pi}{K} ) \big) \text{lr}_0$.} The captions are put in lowercase and all the punctuation characters are erased. We clipp the gradient by l2-norm to 10 to stabilize training and add label smoothing set to 0.1 to the CE loss component. To select our best model among epochs, we used the highest FENSE score instead of using the CE loss on the validation subset. Using FENSE allows you to choose a later training epoch than with the loss of validation function, which gives better results.


\tmp{For the sentence embedding regression method, we used the \texttt{paraphrase-TinyBERT-L6-v2} model\footnote{\url{https://www.sbert.net/docs/pretrained_models.html}}, since it is the one used in the metrics SBERT and FENSE. We also tried larger models (\texttt{all-mpnet-base-v2} and \texttt{all-mpnet-base-v1}), but it does not bring any improvements. The chosen model contains 67M parameters and its weights are frozen during training}. We have set $\beta$ to $1$ in the $\mathcal{L}_s$ function. We tried several values for $\lambda$ (1, 10, 100, 1000, and 10000), and found that 100 is the best parameter for sentence embedding regression. Higher values decrease performance, while lower values have no impact on multitasking compared to the baseline.


\section{Results}

\begin{figure*}[htbp]
    \centering
    \stackunder[5pt]{\includegraphics[width=0.245\linewidth]{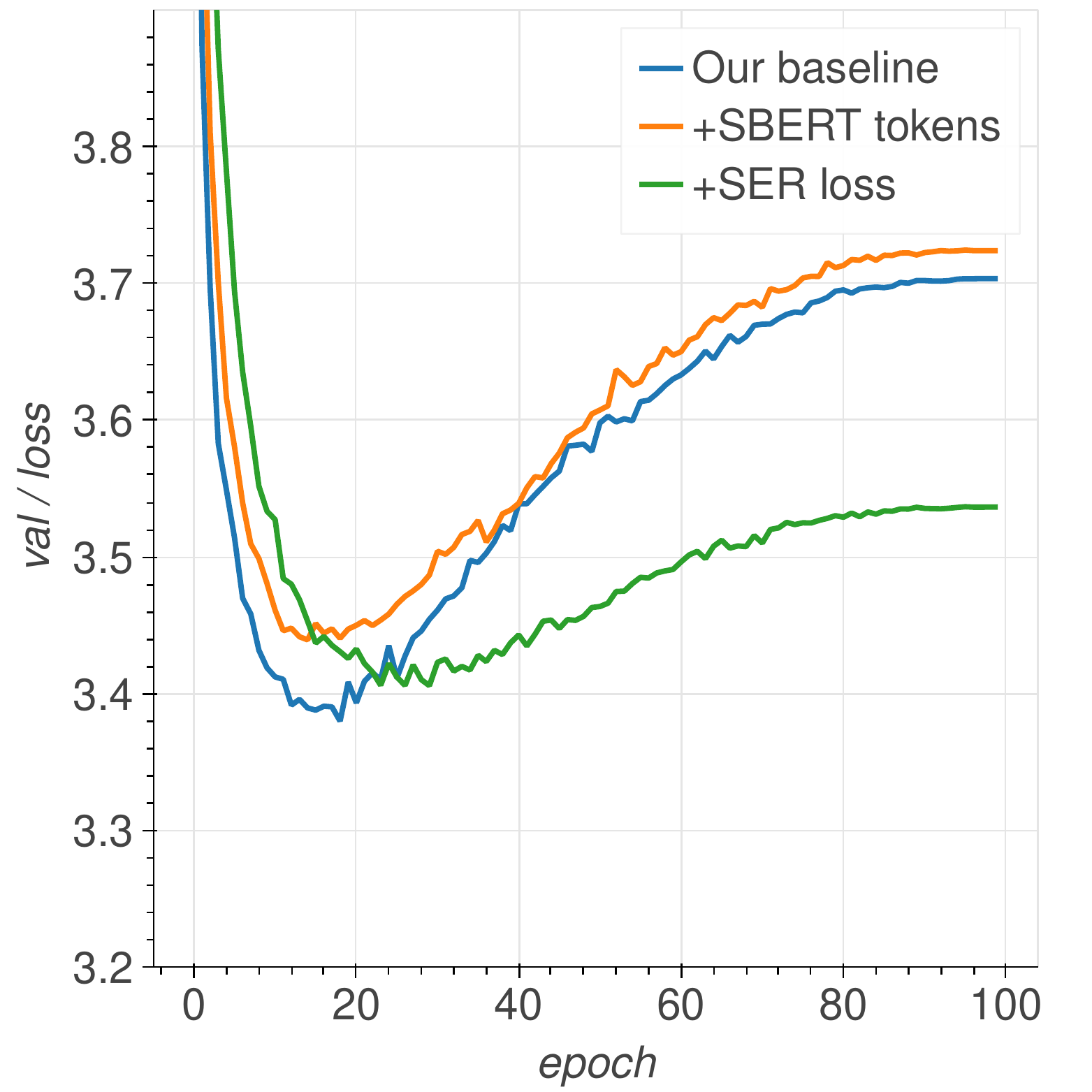}}{\tmp{\footnotesize Loss with wd set to $10^{-6}$.}}
    \stackunder[5pt]{\includegraphics[width=0.245\linewidth]{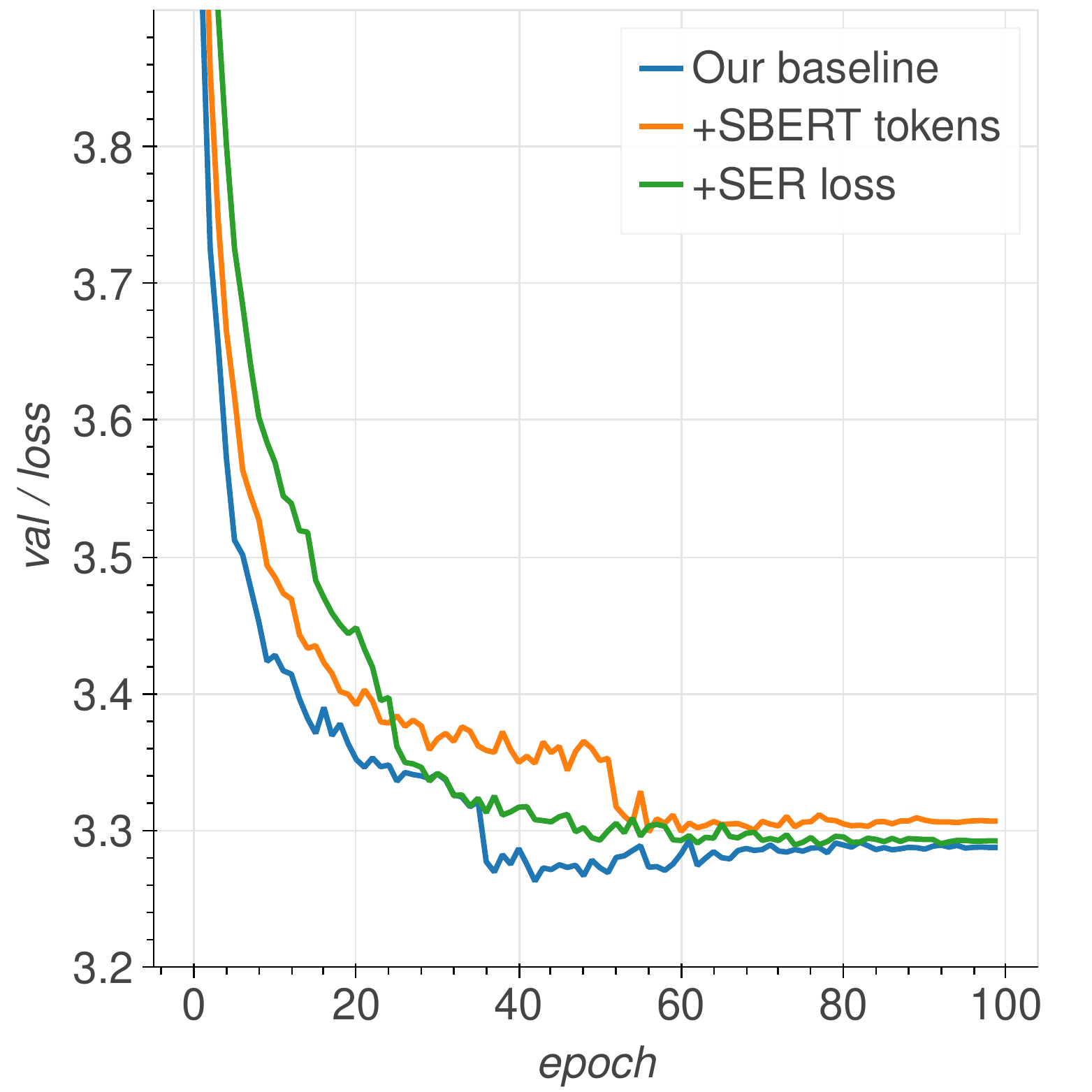}}{\tmp{\footnotesize Loss with wd set to 2.}}
    \stackunder[5pt]{\includegraphics[width=0.245\linewidth]{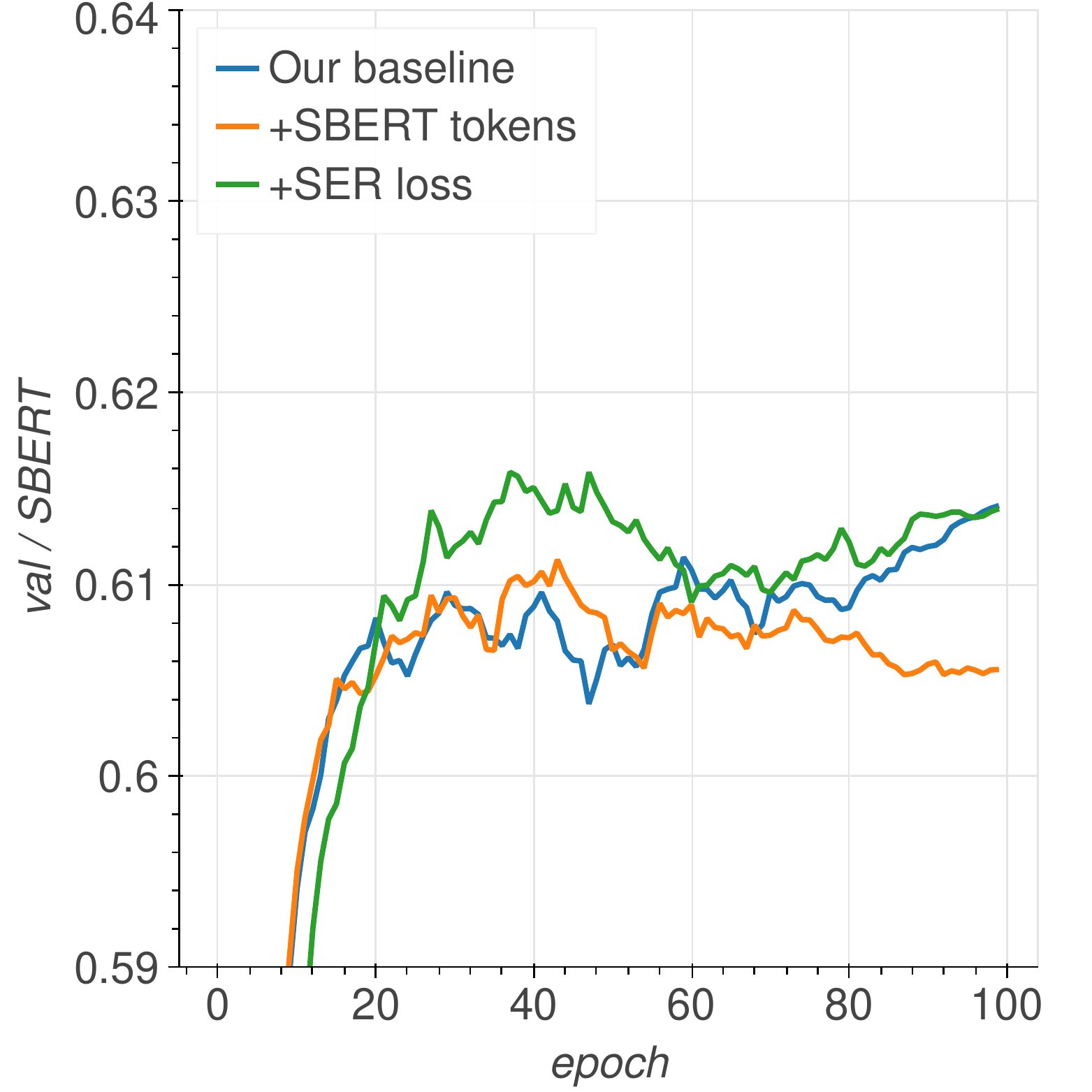}}{\tmp{\footnotesize SBERT with wd set to $10^{-6}$.}}
    \stackunder[5pt]{\includegraphics[width=0.245\linewidth]{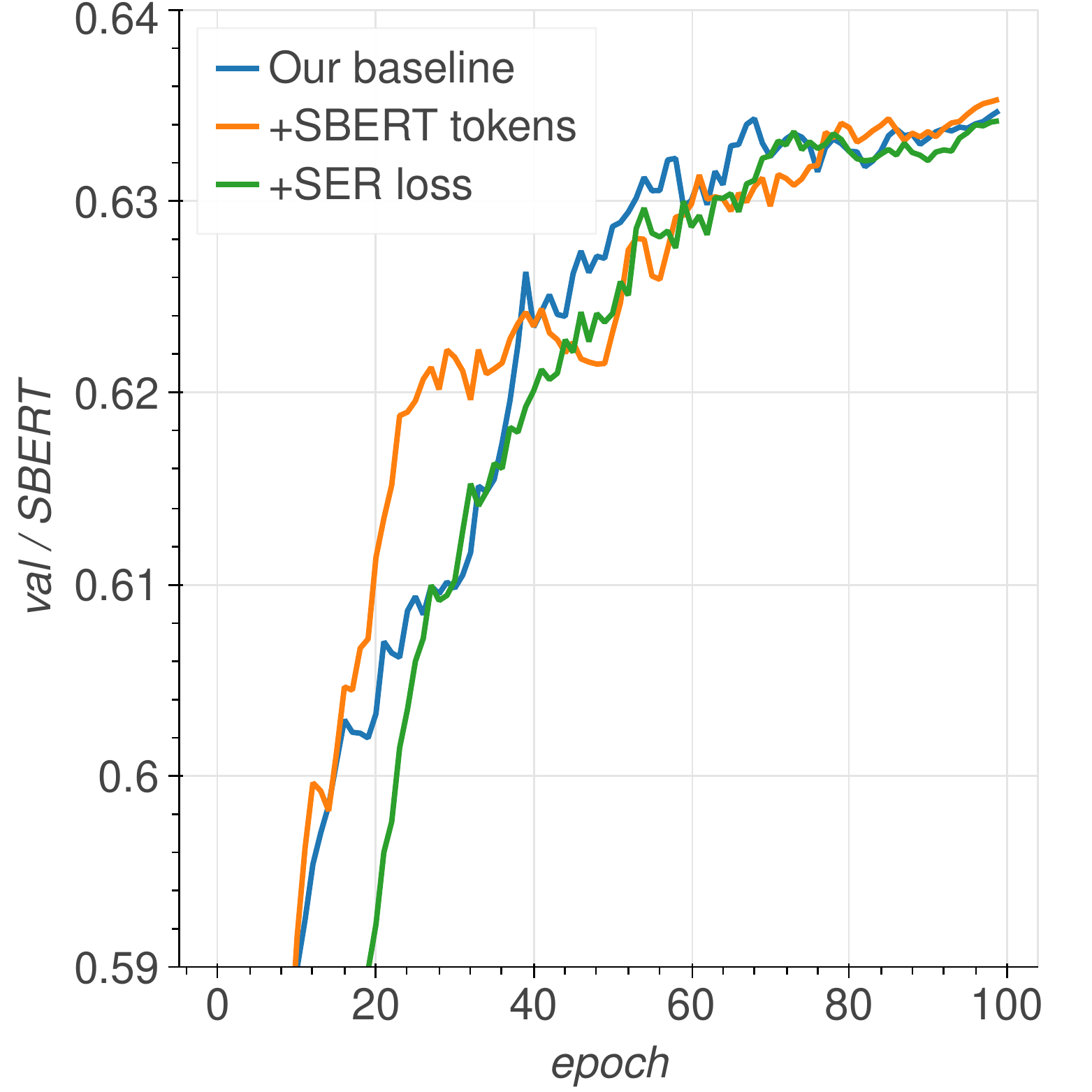}}{\tmp{\footnotesize SBERT with wd set to 2.}}
    \hfill
    \caption{CE losses and SBERT cosine similarities over epochs on validation.}
    \label{fig_val_curves}
\end{figure*}


We reported the scores in table~\ref{tab_results} of our baseline method using a word tokenizer, the method using the SBERT tokenizer (baseline+SBERT tokens) and with the SER loss method (baseline+SBERT tokens+SER loss). All of our scores are averaged over 5 seeds. We also added the other SER method scores named CNN10-trans, the current state-of-the-art scores (Multi-TTA and BYP) and the cross-referencing scores. Cross-referencing is performed by excluding one of the five captions for each audio file and using it as a candidate sentence, while the other four remain the ground truth references. This process is repeated five times to compute an average human agreement score, which we call "cross-references."


\subsection{Discussion}

 \tmp{Using a small wd value, the SER loss shows a slight improvement in FENSE and SPIDEr scores respectively increased from 0.595 to 0.607 and from 0.397 to 0.418}. Moreover, when we introduce a large wd value value to prevent overfitting, we can see a large improvement in FENSE and SPIDEr scores from 0.595 to 0.619 and from 0.397 to 0.445 respectively with our baseline. Nevertheless, even if the SER loss also benefits from the use of a large wd value, the resulting scores became very close to the new baseline which also uses this wd value and the regularization effect given by the SER loss seems no longer significant.
 
 In Fig.~\ref{fig_val_curves}, we can see that the increase in the validation learning curve is reduced by the SER loss with a small wd, which means that it limits the overfitting of our model and explains the gain obtained in the table. The validation losses and SBERT cosine similarities in the figures show that the regularization with a large wd works very well on the model.
 
 The increase in the number of trainable parameters between our baseline and our baseline+SBERT tokens (12.M to 25.7M) comes from the increase in vocabulary (4724 to 30522) which drastically grows the number of parameters in the classifier and the input embedding layer in the decoder part. The method using SBERT tokens and large wd value has become our new best model according to SPIDEr and FENSE, although it is closely followed by our methods using the same decay. \tmp{Our methods are also very close to the current state-of-the-art method which obtain 0.475 compared to our best SPIDEr scores of 0.445 and 0.444, despite the fact that we have eight and four times fewer trainable parameters for our baseline and baseline+SBERT tokens, respectively.}

\subsection{Qualitative analysis}




\tmp{
Tables~\ref{tab_example_286} and \ref{tab_example_269} show several examples of the sentences generated by our model over different training procedures. The baseline system using small wd value seems to try to use more synonyms but fail more often to provide a good description, like in~\ref{tab_example_286}. When using large wd value, the system uses even less words and more generic sentence structures, despite being more accurate. In this case, we also denote that the number of words used in the testing subset decreased from an average of 485.8 words to 387.2 in our baselines systems using small and large wd value, respectively. The reduction is even more important when we add the SER loss, with only 348.8 words used on average.
}

\begin{table}[htb]
    \scriptsize
    \centering
    \caption{Captions for an AudioCaps testing file (id: ``\texttt{\upshape ARFFw0e\_jig}") using the wd value $10^{-6}$ for different methods.}
    \label{tab_example_286}

    \begin{tabular}{|p{3.5cm}|p{1.5cm}|p{1.0cm}|p{1.0cm}|}
        \hline
        \textbf{Candidates} & \textbf{Method} & \textbf{SPIDEr} & \textbf{FENSE} \\ \hline
        a person belching & Our baseline & {.070} & {.208} \\
        a person burps loudly & SBERT tokens & {.135} & {.291} \\
        a person burps loudly several times & SER loss & \first{.252} & \first{.398} \\ \hline
        \multicolumn{4}{|l|}{\textbf{References}} \\ \hline
        \multicolumn{4}{|l|}{loud burping and screaming} \\
        \multicolumn{4}{|l|}{loud burping repeating} \\
        \multicolumn{4}{|l|}{a loud distorted belch followed by a series of burping} \\
        \multicolumn{4}{|l|}{several distorted belches followed by non-distorted burps} \\
        \multicolumn{4}{|l|}{a series of distorted burps followed by non-distorted burps} \\ \hline
    \end{tabular}
\end{table}

\begin{table}[htb]
    \scriptsize
    \centering
    \caption{Captions for an AudioCaps testing file (id: ``\texttt{\upshape JZloTOdIY\_c}") with different weight decays of our baseline.}
    \label{tab_example_269}

    \begin{tabular}{|p{3.5cm}|p{1.5cm}|p{1.0cm}|p{1.0cm}|}
        \hline
        \textbf{Candidates} & \textbf{wd} & \textbf{SPIDEr} & \textbf{FENSE} \\ \hline
        a horse neighs and breathes heavily & $10^{-6}$ & {.284} & \first{.658} \\
        a horse is trotting & $2$ & \first{.337} & {.452} \\ \hline
        \multicolumn{4}{|l|}{\textbf{References}} \\ \hline
        \multicolumn{4}{|l|}{horses growl and clop hooves} \\
        \multicolumn{4}{|l|}{a horse neighs followed by horse trotting and snorting} \\
        \multicolumn{4}{|l|}{horses neighing and snorting while trotting on grass} \\
        \multicolumn{4}{|l|}{horses neighing then snorting and trotting on a dirt surface} \\
        \multicolumn{4}{|l|}{horses neighing and stomping on the ground} \\ \hline
    \end{tabular}
\end{table}

\section{Conclusions}
\tmp{
In this work, we studied the addition of a sentence embedding regression loss component to improve an Automated Audio Captioning system. We searched for the most optimal configuration for our baseline by optimizing hyperparameters, conditioning generation and using a stronger pre-trained encoder. We discovered that the SER loss component seems to limit overfitting for AAC systems, but it does not bring improvement anymore when combined with a stronger regularization method like a large weight decay value. We also noticed that even if the main metrics (SPIDEr, FENSE) are improved by the regularization methods, they do not take into account the diversity of the words used. The diversity of words used could be taken into account in future captioning metrics, or directly by the model during learning like in~\cite{https://doi.org/10.48550/arxiv.1908.04319}.
}


\bibliographystyle{IEEEtran}
\bibliography{refs.bib}

\end{document}